# Enhanced thermopower in a magnetic semiconductor EuTe$_4$ with multiple charge-density-wave instabilities

Hidefumi Takahashi[1,2*], Kiichiro Yoshida[1], Akitoshi Nakano[3] and Shintaro Ishiwata[1,2†]

[1]*Division of Materials Physics and Center for Spintronics Research Network, Graduate School of Engineering Science, The University of Osaka, Osaka 560-8531, Japan*
[2]*Spintronics Research Network Division, Institute for Open and Transdisciplinary Research Initiatives, The University of Osaka, Yamadaoka 2-1, Suita, Osaka, 565-0871, Japan*
[3]*Department of Physics, Nagoya University, Nagoya 464-8602, Japan*

We investigate the layered magnetic semiconductor EuTe$_4$, focusing on its intricate charge-density wave (CDW) states near and above room temperature through single-crystal X-ray diffraction (XRD), magnetic, and thermoelectric measurements. The XRD measurement revealed that the CDW state inducing the polar lattice distortion persists even at 650 K, demonstrating its remarkable thermal stability. Notably, the Seebeck coefficient near room temperature reaches values exceeding 500 µVK$^{-1}$. This large Seebeck coefficient, not fully captured by a simple band calculation, is comparable to those observed in heavy-electron semiconducting oxides, suggesting the importance of electron correlation and spin/lattice instabilities. Furthermore, potentially reflecting the competition of two types of CDW states, the thermal conductivity near room temperature is as low as 0.02 Wcm$^{-1}$K$^{-1}$. As a result, the thermoelectric figure of merit $zT$ reaches 0.22 at 460 K. These findings establish EuTe$_4$ as a compelling platform to explore novel types of thermoelectric materials with multiple electronic instability.

*takahashi.hidefumi.es@osaka-u.ac.jp
†ishiwata.shintaro.es@osaka-u.ac.jp

## I. INTRODUCTION

Low-dimensional materials exhibiting charge-density wave (CDW) states have garnered considerable interest due to their rich interplay of electronic and structural properties, as well as their potential for advanced device applications[1]. Among these systems, transition-metal chalcogenides such as 1T-TaS$_2$, 1T-TiSe$_2$, and NbSe$_3$ have been studied as prototypical platforms for investigating CDW physics[1–5]. These materials exhibit a variety of intriguing phenomena, including metal-insulator transitions, nonlinear transport, and collective electronic behaviors driven by strong electron-lattice coupling. Furthermore, their tunability via external stimuli such as temperature, pressure, and chemical doping has opened avenues for potential applications in memory devices and sensors.[6–10].

Recently, rare-earth tritellurides (RTe$_3$) have been extensively studied because of rich CDW states[11–13], which can be tuned through external perturbations such as chemical intercalation and applied pressure[14]. Importantly, increasing the atomic number of the rare-earth element in RTe$_3$ modifies the Fermi surface nesting, leading to the emergence of a second CDW state[15]. The coexistence or competition between distinct CDW states enriches the phase diagram, which includes a superconducting phase[16,17]. Additionally, the rare-earth elements induce complex spin textures through global and local mirror-symmetry breaking associated with the CDW state[15,18].

A europium tetratelluride EuTe$_4$ also exhibits unique CDW orders as shown in Fig. 1(a). Unlike its tritelluride counterparts, where partial Fermi surface nesting allows metallic conduction, EuTe$_4$ exhibits a fully gapped Fermi surface, rendering it semiconducting below the CDW transition temperature[19–22]. XRD measurements have confirmed the incommensurate nature of the CDW, characterized by a modulation vector $q$~$0.67b^*$, while some experiments suggest the presence of multiple CDW states[22–26]. One of the remarkable features of EuTe$_4$ is an exceptionally large thermal hysteresis in a temperature range including room temperature, likely arising from the complexity of the CDW orders. Consequently, EuTe$_4$ demonstrates non-volatile, optically and electrically tunable polar order in the CDW state at room temperature, highlighting its potential for applications in ultrafast electronic and photonic devices[27,28]. In addition, the magnetic Eu$^{2+}$ ions provide an opportunity to study the coupling between the complex CDW states and magnetism[29,30].

Here, we report on the structural, magnetic, and thermoelectric properties of EuTe$_4$ single crystals. XRD experiments reveal that the CDW state in EuTe$_4$ persists even above 400 K. Magnetic measurements indicate significant in-plane anisotropy, which reflects the lattice distortion with the polar lattice displacement along the *a* axis. Both electrical resistivity and thermoelectric power exhibit pronounced hysteresis over a broad temperature range, with the thermoelectric power reaching an exceptional value of 500 μVK$^{-1}$ near and above room temperature. This behavior cannot be fully captured by simplified first-principles calculations, suggesting the pivotal role of many-body effects, including electron-electron interactions and magnetism. These results establish EuTe$_4$ as a compelling platform for studying the interplay between CDW order, magnetism, and thermoelectric phenomena.

## II. METHOD
### A. Crystal growth and Transport measurement

Single crystals of EuTe$_4$ were grown using the Te flux method[19]. Eu lumps (99.9 %) and Te granules (99.9999 %) were mixed with a ratio of 1 : 15. The total weighted starting materials were sealed in an evacuated fused silica tube in vacuum followed by heating at 850 °C for 2 days in a muffle furnace. Then, the furnace was slowly cooled to 430 °C in 24 h, and held at this temperature for two weeks then decanted using a centrifuge. The large single crystal shown at the bottom of Fig.1(a) was used for the transport measurements. The crystal structure was identified by the synchrotron XRD measurements for single crystalline samples at BL02B1 in SPring-8. Magnetic properties, electrical resistivity, and Seebeck coefficient (thermopower) below 400 K were measured using a magnetic property measurement system (MPMS, Quantum Design, Inc.). The resistivity was measured by a standard four-terminal method. Resistivity between 290 and 490 K was measured on a hot plate in air. The Seebeck coefficient was measured by a conventional steady-state method. Thermal conductivity was measured by a conventional four-terminal steady-state method using a He closed-cycle refrigerator (from 10 to 470 K). The Seebeck coefficient was measured between 250 and 460 K using the same refrigerator. Due to limitations in the temperature-control stability of the refrigerator, reliable data could be acquired only on the warming branch for $T \lesssim 300$ K and on the cooling branch for $T \gtrsim 300$ K; consequently, the Seebeck coefficient and thermal conductivity measured using the refrigerator were shown only for the warming run below ~300 K and the cooling run above ~300 K. The composition of single crystals was determined by energy dispersive X-ray (EDX) spectroscopy, a JEOL JCM-7000 scanning electron microscope (SEM). EDX measurements were performed at different positions on the crystal surfaces within an instrument error of 1−2% (for details, see Fig. S1 and Table S1, respectively, in the Supplemental Material[31]). The Eu element tends to be deficient by less than 10% from the nominal composition (Eu:Te = 1:4). Thermogravimetry–differential thermal analysis (TG–DTA, STA-2500 Regulus made by NETZSCH Company) were performed with 8.2 mg of crushed single crystals in Al$_2$O$_3$ capsules.

### B. Band calculations

The non-relativistic bulk electronic structure calculations of EuTe$_4$ were performed within the density functional theory (DFT) using the Perdew-Burke-Ernzerhof (PBE) exchange-correlation functional as implemented in the Quantum ESPRESSO code[32–35]. The projector augmented wave (PAW) method has been used to account for the treatment of core electrons[35]. The cut-off energy for plane waves forming the basis set was set to 50 Ry. The lattice parameters and atomic positions were taken from the single-crystal XRD experiment. The corresponding Brillouin zone was sampled using a 11 × 11 × 5 $k$-mesh to calculate the PWscf, and a 30 × 30 × 30 $k$-mesh for normal states and 10 × 20 × 30 $k$-mesh for CDW-I states to perform Fermi surface visualization by the FermiSurfer package[36]. We neglect the Eu $f$ orbitals since they are predicted to be more than 1.5 eV below the Fermi surface[37]. The Seebeck coefficient was calculated using the BoltzTraP[38].

## III. RESULTS AND DISCUSSION
### A. Structural analysis and magnetic properties

Figure 1 (c) exhibits the XRD patterns for (0$KL$) plane and Table 1 summarizes the structure parameters at 650 K (the XRD pattern for ($HK$0) plane at 650 K is shown in Fig. S2[31]). The analyzed structure is shown in Fig. 1(a) as CDW-I. It is noted that the space group of the CDW-I phase, $P2_1cn$ (No. 33), is a polar structure with no inversion symmetry. While previous studies have suggested a transition from the CDW state to the normal state near room temperature [19], the present measurement confirmed a nearly commensurate superlattice peak around (1, ±1/3, 0) corresponding to the previously reported 3-fold CDW state even at 650 K, as denoted by the green arrows. This result indicates the realization of the nearly 3-fold CDW state at far above room temperature. Although the modulation is incommensurate, we performed structural refinement assuming a commensurate 3-fold modulation along the $b$ axis as an average structure, which effectively captures the lattice distortion associated with the high-temperature CDW phase. Some previous studies also suggest that CDW transition occurs above 600 K[24], and the incommensurate modulation wave vector nearly $q \sim 0.67b^*$ persists at 300 K[20,21,23,36]. Angle-resolved photoemission spectroscopy (ARPES) measurements also show that the CDW gap is already open at around 400 K or even higher[20]. We have also measured XRD patterns below 400 K and obtained complex diffraction patterns, reflecting the influence of different CDW phases[22,24,25] or surface domains[20]. Therefore, the precise structural analysis has not been achieved below 400 K. Figure 1(c) compares the single-crystal X-ray diffraction patterns for the (0$KL$) plane recorded at 650 K and 100 K. The superlattice reflection corresponding to CDW-I indicated by the green arrow at 650 K becomes substantially more intricate at 100 K, most likely owing to domain formation and the coexistence of competing CDW orders; in addition, a new, well-defined reflection appears within the region delineated by the red rectangle. Figure 1(d) shows an enlarged view of this region and the temperature evolution of the associated superlattice peak at $q \sim (0, K, L) + (0, 0.04, 0.5)$. Upon both cooling and subsequent heating, this peak develops a pronounced hysteresis that mirrors the hysteresis observed in the temperature dependences of the electrical resistivity and Seebeck coefficient discussed below. These observations indicate that the lattice modulation in the CDW-II phase can be associated with this superlattice reflection.

Figure 1(b) illustrates a schematic diagram of the temperature evolution of the CDW states. The high-temperature CDW-I state [$q_{CDW} \sim (0, 1/3, 0)$] changes to the low-temperature CDW-II state [$q_{CDW} \sim (0, 0.04, 0.5)$] at around 300 K. Here, we define the CDW modulation vectors $q_{CDW}$ with respect to the unit cell of the normal state shown in Fig. 1(a). Previous study suggests the possibility of a transition between the normal state and the CDW-I state at around 726 K on heating and 652 K on cooling [24]. To determine the stability of the crystals at high temperatures, we performed thermogravimetric differential thermal analysis (TG-DTA) and measured XRD before and after the process as shown in Fig. S3[31]. The results revealed that the crystals decomposed at around 700 K during the heating process, and an anomaly was observed at around 660 K during the cooling process, probably due to crystallization into Eu$_3$Te$_7$ [39]. In addition, the normal state structure could not be observed even with the single crystal XRD measurements at temperatures above 650 K.

Figure 2(a) shows the temperature dependence of the magnetic susceptibility χ for $H//a$, $H//b$, and $H//c$ at $H$ = 0.1 T. The compound undergoes an antiferromagnetic (AFM) transition at 7 K ($T_N$) as observed in previous study[19]. The magnetic susceptibility in $H//a$ is significantly reduced below $T_N$ as compared to that in $H//b$ and $H//c$ [20], indicating that spin moments are parallel or antiparallel to the $a$-axis in the AFM phase. The spin anisotropy within the $ab$ plane in the AFM phase of EuTe$_4$ can be associated with the orthorhombic distortion caused by the CDW transition. In the case of the similar layered compound EuTe$_2$, the spins are oriented along the $c$-axis, consistent with the tetragonal symmetry[40]. In addition, the spin orientation in EuTe$_4$ can be influenced by the Dzyaloshinskii-Moriya (DM) interactions resulting from an inversion symmetry breaking with polarity along the $a$-axis. The AFM nature is also confirmed by the Curie–Weiss fits for $H//a$ in the range between 10 K and 350 K, as shown in Fig. S4 (see Supplemental Material[31]), which resulted in Weiss temperature of $\theta_W$ = -2.5 K. The estimated effective magnetic moment, 7.32 $\mu_B$, is close to the theoretical value for EuTe$_4$ with Eu$^{2+}$ (7.94 $\mu_B$), as previously confirmed by the X-ray absorption experiment[23]. Here we assume that the valence of the Te ions in the Eu-Te slab is -2, and that the Te monolayers and bilayers are neutral[23].

To further investigate the anisotropy in the magnetically ordered states, we measured the magnetic-field dependence of the magnetization $M$ for $H//a$, $H//b$, and $H//c$ at 1.8 K as shown in Fig. 2(b). For $H//b$ and $H//c$, the magnetization curves are linear to the field and the former shows saturation at approximately 5.7 T. On the other hand, the magnetization curve for $H//a$ exhibits sudden jump at approximately 2 T, which implies a spin-flop transition, and shows saturation at 4.5 T. The saturation field for $H//a$ is lower than the others, indicating that the localized moment tends to align along the $a$-axis direction as suggested by the temperature dependent magnetic susceptibility below $T_N$. Furthermore, although the magnetization exhibits negligible hysteresis over the entire field range, the magnetoresistance displays a pronounced hysteresis loop (see Fig. S5 in the Supplemental Material[31]). We attribute this discrepancy to a field-induced subtle lattice distortion by the spin reconfiguration, that leaves the net magnetization almost unchanged yet produces the hysteresis of magnetoresistance, potentially associated with the multidomain state discussed below. These observations provide direct evidence for strong coupling among the spin, lattice, and charge modulation.

**B. Thermoelectric properties**

Figure 3(a) exhibits the temperature dependence of the electrical resistivity $\rho$ for $I//a$ and $I//b$. A hysteresis is observed over a wide temperature range between 200 K and 400 K measured using MPMS, consistent with prior studies[19,20]. We repeated the measurement several times and consistently reproduced the same loop, indicating that extrinsic factors such as cracks in the crystal have a negligible influence on the result. The broad hysteresis loop suggests the emergence of distinct CDW states at lower temperature. Furthermore, the hysteresis loop remains unclosed even when increasing the highest measured temperature to 490 K by using a hot plate in air, suggesting that the low-temperature CDW-II phase partially remains up to rather high temperatures, where the CDW-I is expected to be stable[23]. On the other hand, the small hysteretic behavior can be found around 100-200 K, indicating persistent effects of CDW domains at lower temperatures. In contrast to previous reports, we do not observe a pronounced increase in electrical resistivity at low temperatures characteristic of a semiconducting electronic structure. This discrepancy may arise from the different impurity levels in the Eu source: we used 99.9 % pure Eu, whereas the earlier studies employed 99.999 % pure Eu. A 0.1 % deviation in stoichiometry results in a change in carrier density of roughly $10^{18}$ to $10^{19}$ cm$^{-3}$, which is large enough to affect the low-temperature transport of a narrow-gap semiconductors.[41] EDX measurements also suggest a few percent deficiency of Eu, which introduces hole carriers to the system. While resistivity exhibits minimal in-plane anisotropy near room temperature, a slight anisotropy is found at low temperatures, potentially reflecting a quasi-one-dimensional band structure, as previously indicated by ARPES measurements[20–22]. Additionally, a subtle anomaly near 70 K along the

$I/\!/a$ orientation may correspond to the emergence of another distinct CDW phase, as recently suggested in scanning tunneling microscopy (STM) studies[25].

To elucidate the impact of CDWs on the electronic properties, we investigated the temperature dependence of thermoelectric properties, as shown in Fig. 3(b). Whereas a similar pronounced hysteresis is observed over a broad temperature range, the hysteresis loop closes around 230 K in the temperature dependence of Seebeck coefficient. Since Seebeck coefficient is less sensitive to electron scattering as compared to electrical resistivity, the thermal hysteresis below 230 K in the resistivity can be ascribed to the domain dynamics and the large thermal hysteresis around 300 K in Seebeck coefficient suggests a first-order transition between distinct electronic states with different CDW modulations; i.e., high-temperature CDW-I phase and low-temperature CDW-II phase. While previous studies have indicated the presence of multiple CDWs, the detailed structural nature of these phases remains unresolved.

Remarkably, the Seebeck coefficient at room temperature (~300 K) reaches exceptionally high values of 550 $\mu$VK$^{-1}$ and 400 $\mu$VK$^{-1}$ during heating and cooling processes, respectively. In the temperature range from room temperature to 460 K, where the large Seebeck coefficient is observed, electrical resistivity becomes as low as approximately 0.02 $\Omega$ cm during the cooling process. Consequently, the present compound exhibits a relatively high power factor (PF = $S^2/\rho$), reaching up to 10 $\mu$W K$^{-2}$cm$^{-1}$ at 460 K as shown in Fig. S6 (see Supplemental Material[31]). This value is comparable to that of La(Fe,Co)$_4$Sb$_{12}$, a well-known thermoelectric material, which exhibits PF of ~6.66 $\mu$W K$^{-2}$cm$^{-1}$ at 300 K[42]. Such high PF suggests that EuTe$_4$ has potential as an efficient thermoelectric material in the temperature range above room temperature. To evaluate the dimensionless figure of merit ($zT$), the temperature dependence of the thermal conductivity was measured as shown in Fig 3(c). The thermal conductivity is relatively low, approximately 0.01 Wcm$^{-1}$K$^{-1}$ at 300 K, especially for the temperature gradient along the $a$ axis, and the peaky structure typically observed for clean semiconductors is absent as in the case of a glass. In sharp contrast, the rare-earth tritellurides RTe$_3$ (R = Gd, Tb, Dy, Ho, Er) with CDW orders display a much larger lattice thermal conductivity of 0.04–0.08 Wcm$^{-1}$K$^{-1}$ (total $\kappa$ = 0.16–0.25 Wcm$^{-1}$K$^{-1}$) and the conventional low-temperature upturn characteristic of crystalline solids[43]. The glass-like behavior found in EuTe$_4$ is therefore most plausibly attributed to a multidomain state generated by the competition of nearly degenerate CDW orders. This scenario is supported by the broad hysteresis observed in the temperature dependence of the resistivity and by STM measurements that reveal pronounced CDW competition and domain formation at low temperatures[20,25].

The temperature dependence of $zT$ for the $a$-axis direction is presented in Fig. 3(d). The $zT$ increases significantly above 250 K, reflecting the rapid decrease in electrical resistivity, and reaches a value of 0.22 at 460 K. This value is comparable to that of filled skutterudite antimonides and transition metal chalcogenide 1T-TiS$_2$ [42,44]. One of the key factors contributing to the observed high thermoelectric performance is the large thermopower of the system. This significantly large value (~ 500 $\mu$VK$^{-1}$) suggests the presence of a gap-like electronic structure, as reported in ARPES studies[20], which is inconsistent with the first-principles band calculations as shown in Fig. S8 (Supplemental Material). This large Seebeck coefficient is comparable to that of CoSb$_3$ (~450 $\mu$VK$^{-1}$), studied as a promising thermoelectric material[45], where the large Seebeck coefficient is attributed to a low carrier concentration of approximately 4×10$^{17}$ cm$^{-3}$. However, as shown in Fig. 4(a), the carrier concentration $n$ in EuTe$_4$, derived from Hall resistivity measurements (see Fig. S7 in the Supplemental Material[31]), reveals a much higher carrier concentration of ~10$^{19}$ cm$^{-3}$ at 300 K. For materials with similar carrier concentrations, typical Seebeck coefficients range from 200 to 300 $\mu$VK$^{-1}$[46], suggesting that the large Seebeck coefficient of EuTe$_4$ cannot be explained by the conventional one-electron approximation.

In nondegenerate semiconductors with the carrier concentration less than 10$^{20}$ cm$^{-3}$, the Seebeck coefficient $S$ is generally expressed as a logarithmic function of carrier concentration, as $S \sim \ln(n^{-1})$[47]. The logarithmic function in Ref.[47] arises from the dependence of both the Seebeck coefficient and the carrier concentration on the energy gap $\Delta$ in nondegenerate semiconductors, where the Seebeck coefficient scales as $S \sim \Delta/k_BT$, and the carrier concentration $n$ follows an exponential behavior, $n \sim \exp(-\Delta/k_BT)$. Notably, the $S-\ln(n^{-1})$ trend observed in EuTe$_4$ deviates significantly from that of a typical semiconductor, Ge, shifting towards higher $S$ values as shown in Fig. 4(b). This deviation suggests additional contributions such as increased electron mass, multiple Fermi pockets, magnetism, and structural fluctuations[48–56]. In particular, a moiré-like CDW structure has been proposed as the driver of the emergent semiconducting band structure[26]. If the observed deviations of the $S-\ln(n^{-1})$ trend from that of Ge were caused solely by an enhanced carrier mass, the effective mass would need to be an order of magnitude larger than the free-electron mass.

To understand the origin of this deviation, we compare EuTe$_4$ with other semiconductors exhibiting heavy carrier mass at room temperature such as the SrTiO$_3$ system ($m^* \sim 9$–$13$ $m_0$)[47], several half-Heusler compounds ($m^* \sim 4$–$6$ $m_0$)[57,58], and the (Cu, Li)-doped MnTe system ($m^* \sim 2$–$7$ $m_0$).[59,60] These compounds fall along nearly the same straight line in Fig. 4(b), and their power factors lie in the same range (1–30 μW K$^{-2}$cm$^{-1}$) as that of EuTe$_4$. Furthermore, the Fermi surface obtained from first-principles calculations in the CDW-I phase along with the band dispersions in Figs. S8 and S10[31], reveals an essentially two-dimensional, multi-pocket band structure. ARPES confirms that EuTe$_4$ is semiconducting, with no Fermi surface at $E_F$; nevertheless, at an energy of about –0.2 eV the spectra display a two-dimensional, multi-pocket structure that closely matches the calculated Fermi surface[20,21]. This characteristic band structure just below $E_F$ may therefore be crucial for producing the unusually large Seebeck coefficient of EuTe$_4$. In the case of MnTe, the power factor is likely enhanced further by paramagnon drag far above Néel temperature ($T_N \sim 315$ K), reflecting the persistence of short-range antiferromagnetic correlations up to $\sim600$ K. Because EuTe$_4$ orders antiferromagnetically only below $T_N \sim 7$ K, a comparable paramagnon contribution near room temperature is unlikely; instead, the material has competing CDW instabilities well above room temperature, and thus, a phason-drag mechanism—the CDW analogue of magnon drag—may boost the Seebeck coefficient.[61]. To confirm these possibilities as the origin of the exceptionally large thermopower of EuTe$_4$, a thorough investigation of this competing CDW states is essential.

## IV. CONCLUSIONS

In conclusion, we have successfully synthesized single crystals of EuTe$_4$ and conducted comprehensive investigations of its structural, magnetic, and thermoelectric properties. Structural analysis revealed that the crystal exhibits a CDW state at as high as 650 K. Measurements of the magnetic and thermoelectric properties revealed in-plane anisotropy, associated with the polar lattice distortion and a large $zT$ value of 0.22 at 460 K, reflecting a large Seebeck coefficient of 550 μVK$^{-1}$ and a low thermal conductivity of 0.02 Wcm$^{-1}$K$^{-1}$. Band structure calculations based on the CDW structure failed to reproduce the semiconducting gap and the observed large Seebeck coefficient. This discrepancy underscores the need for further studies on EuTe$_4$, including detailed structural investigations of the CDW state and theoretical approaches incorporating many-body effects, such as magnetic interactions and electron-electron correlations, to fully elucidate the enhanced thermoelectric performance.

## ACKNOWLEDGMENTS
The authors thank T. Nomoto for fruitful discussions about band calculations. This study was supported in part by KAKENHI (Grant No. JP23H04871, JP23K13059, JP24H01621, JP24K00570, JP25H00420, and JP25H01248), FOREST (No. JPMJFR236K) from JST, Murata Foundation, Yazaki Memorial Foundation for Science, Technology, and Asahi Glass Foundation.

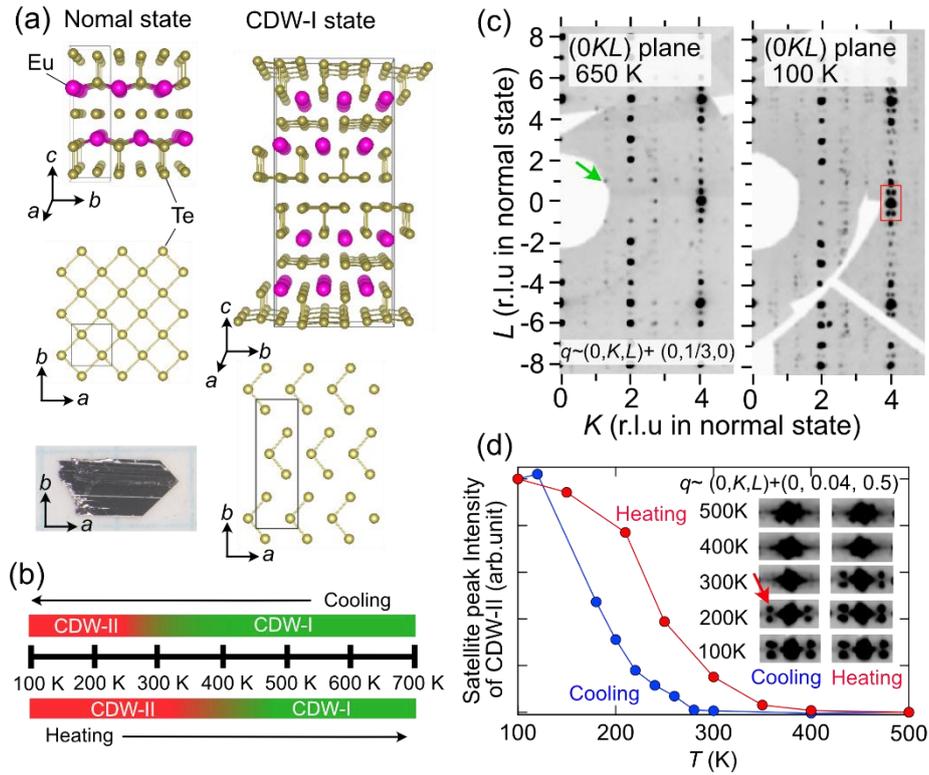

**Fig. 1** (a) Crystal structures of EuTe$_4$ of normal state and CDW state reported in Ref. [19]. (b) Schematics of the temperature-dependent stability of CDW states. (c) Single-crystal X-ray diffraction patterns for the (0$KL$) plane measured at 650 K (left panel) and 100 K (right panel). The green arrow identifies the CDW-I superlattice reflection, whereas a new peak ascribed to the CDW-II phase appears within the red-boxed region. (d) Enlarged view of the red-boxed region in Fig. 1(c) and the temperature dependence of the CDW-II superlattice reflection at $q \sim (0, 4\pm0.04, 0.5)$.

**Table 1** Refined structural parameters for EuTe$_4$ at 650 K determined by single-crystal synchrotron X-ray diffraction. The space group is $P2_1cn$, and the obtained lattice parameters are $a$ = 4.5861(3) Å, $b$ = 13.9979(11) Å, $c$ = 31.653(13) Å. The number of unique reflections for the refinement within the resolution limit $d > 0.5$ Å is 4774. The obtained $R$ factor and GOF are 6.87% and 1.017, respectively.

| | Site | x | y | z | Occ. | $U_{iso}$ (Å$^2$) |
|---|---|---|---|---|---|---|
| Eu1 | 4a | 0.5794(4) | -0.74725(7) | 0.16522(3) | 1 | 0.0392(2) |
| Eu2 | 4a | -0.4215(4) | -0.41333(5) | 0.16603(4) | 1 | 0.0390(3) |
| Eu3 | 4a | 0.0720(4) | 0.08217(6) | -0.16550(4) | 1 | 0.0399(3) |
| Te1 | 4a | 0.0317(10) | 0.07782(8) | -0.05331(5) | 1 | 0.0490(6) |
| Te2 | 4a | 0.0603(9) | -0.25063(12) | -0.05371(4) | 1 | 0.0552(4) |
| Te3 | 4a | 0.6303(9) | -0.41740(7) | 0.05376(5) | 1 | 0.0415(4) |
| Te4 | 4a | 0.5778(6) | -0.08540(6) | -0.14340(5) | 1 | 0.0359(3) |
| Te5 | 4a | 0.5775(8) | -0.24707(9) | 0.25046(3) | 1 | 0.0519(4) |
| Te6 | 4a | 0.0880(8) | -0.41452(7) | 0.25013(5) | 1 | 0.0504(6) |
| Te7 | 4a | 0.0768(6) | -0.24750(9) | 0.14213(3) | 1 | 0.0362(2) |
| Te8 | 4a | 0.0784(5) | -0.58053(6) | 0.14320(5) | 1 | 0.0350(3) |
| Te9 | 4a | 0.5680(9) | -0.58143(8) | 0.25013(5) | 1 | 0.0526(5) |
| Te10 | 4a | 0.0324(10) | -0.57764(8) | 0.05485(5) | 1 | 0.0492(5) |
| Te11 | 4a | 0.6310(10) | -0.08181(8) | -0.05497(5) | 1 | 0.0415(4) |
| Te12 | 4a | 0.0613(9) | -0.24929(11) | 0.05289(4) | 1 | 0.0518(4) |

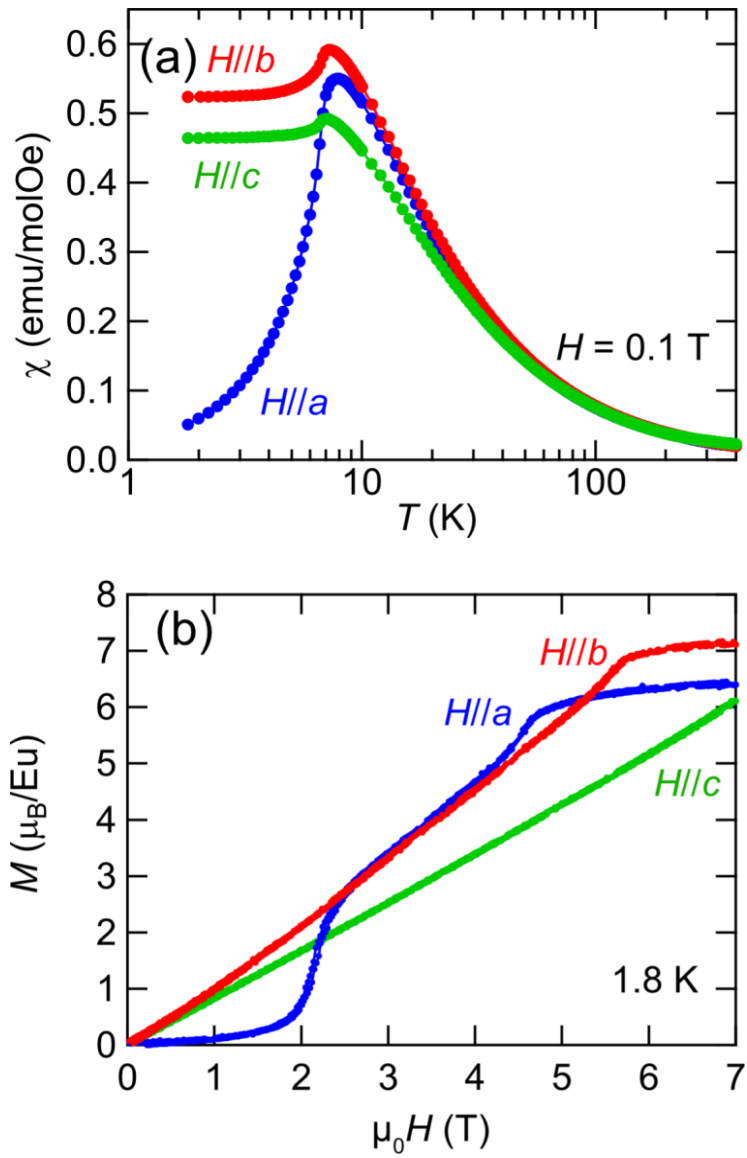

**Fig. 2** (b) Temperature dependence of the magnetic susceptibility measured on zero field cooling (ZFC) for *H//a*, *H//b* and *H//c* at 0.1 T. (b) Magnetization as a function of the magnetic field for *H//a*, *H//b* and *H//c* at 1.8 K, where negligible hysteresis is observed for all field orientations.

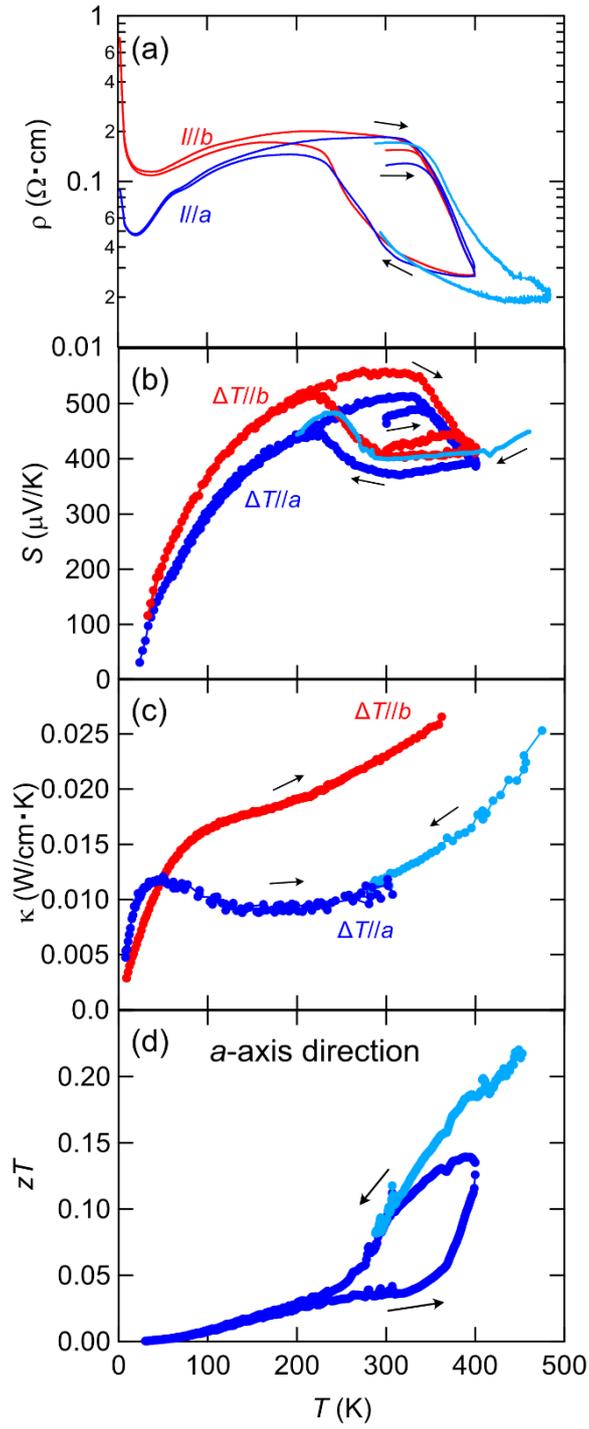

**Fig. 3** Temperature dependence of (a) electrical resistivity for *I*//*a* and *I*//*b*, (b) Seebeck coefficient for Δ*T*//*a* and Δ*T*//*b*, (c) thermoelectric conductivity for Δ*T*//*a* and Δ*T*//*b*, (d) dimensionless figure of merit (*zT*) for *I*//*a* and Δ*T*//*a*.

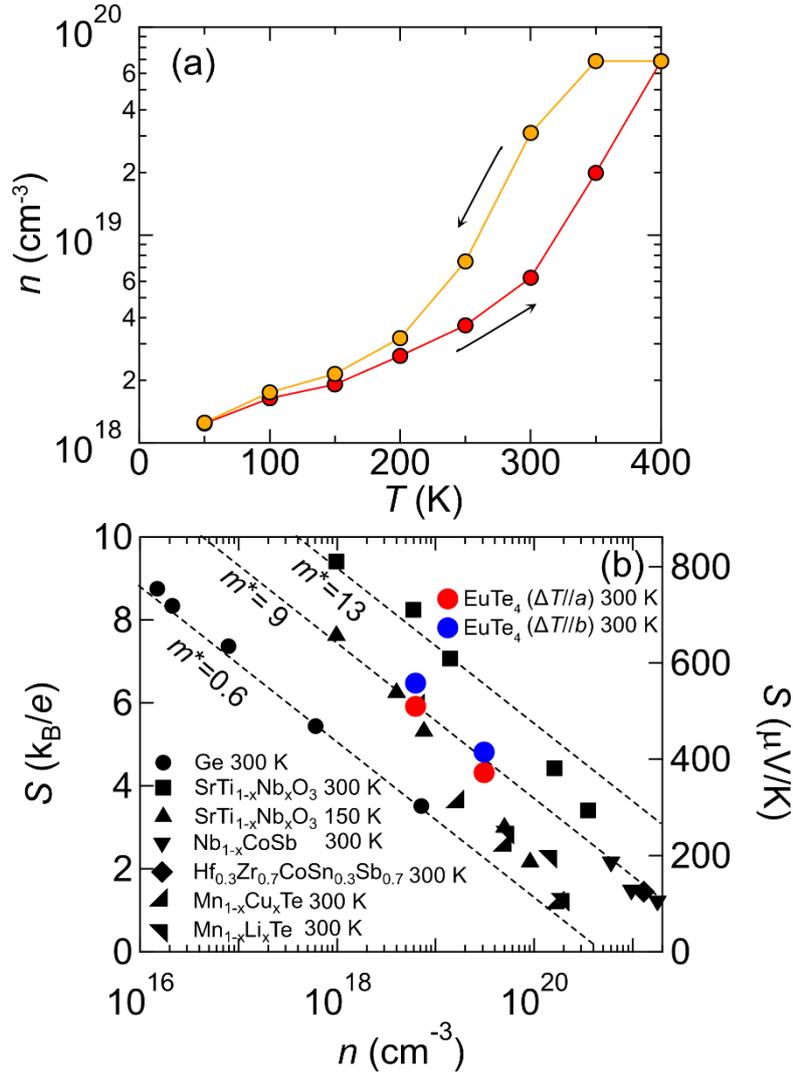

**Fig. 4** (a) Temperature dependence of carrier concentrations $n$ estimated from the Hall resistivity ($I//a$, $V//b$, $H//c$). (b) The Seebeck coefficient $S$ of EuTe$_4$, $p$-type Ge, Nb-doped strontium titanate, half-Heusler compounds, and (Cu, Li)-doped MnTe as a function of carrier concentration $n$ at selected temperatures[46, 57-60]. In all systems, $S$ follows a $\ln(n^{-1})$ dependence, as expected by Eq. (4) in Ref. [46].

**REFERENCES**

[1] G. Grüner, The dynamics of charge-density waves, Rev. Mod. Phys. **60**, 1129 (1988).
[2] K. Nakanishi, H. Takatera, Y. Yamada, and H. Shiba, The nearly commensurate phase and effect of harmonics on the successive phase transition in 1T-TaS$_2$, J. Phys. Soc. Jpn. **43**, 1509 (1977).
[3] S. Tanda, T. Sambongi, T. Tani, and S. Tanaka, X-Ray Study of Charge Density Wave Structure in 1T-TaS$_2$, J. Phys. Soc. Jpn **53**, 476 (1984).
[4] R. E. Thomson, B. Burk, A. Zettl, and J. Clarke, Scanning tunneling microscopy of the charge-density-wave structure in 1T-TaS$_2$, Phys. Rev. B **49**, 16899 (1994).

[5] G. Grüner, A. Zawadowski, and P. M. Chaikin, Nonlinear Conductivity and Noise due to Charge-Density-Wave Depinning in $NbSe_3$, Phys. Rev. Lett. **46**, 511 (1981).

[6] G. Liu, B. Debnath, T. R. Pope, T. T. Salguero, R. K. Lake, and A. A. Balandin, A charge-density-wave oscillator based on an integrated tantalum disulfide-boron nitride-graphene device operating at room temperature, Nat. Nanotechnol. **11**, 845 (2016).

[7] A. A. Balandin, S. V. Zaitsev-Zotov, and G. Grüner, Charge-density-wave quantum materials and devices—New developments and future prospects, Appl. Phys. Lett. **119**, 170401 (2021).

[8] G. Liu, E. X. Zhang, C. D. Liang, M. A. Bloodgood, T. T. Salguero, D. M. Fleetwood, and A. A. Balandin, Total-ionizing-dose effects on threshold switching in $1T-TaS_2$ charge density wave devices, IEEE Electron Device Lett. **38**, 1724 (2017).

[9] A. Khitun, G. Liu, and A. A. Balandin, Two-dimensional oscillatory neural network based on room-temperature charge-density-wave devices, IEEE Trans. Nanotechnol. **16**, 860 (2017).

[10] A. G. Khitun, A. K. Geremew, and A. A. Balandin, Transistor-less logic circuits implemented with 2-D charge density wave devices, IEEE Electron Device Lett. **39**, 1449 (2018).

[11] E. DiMasi, M. C. Aronson, J. F. Mansfield, B. Foran, and S. Lee, Chemical pressure and charge-density waves in rare-earth tritellurides, Phys. Rev. B **52**, 14516 (1995).

[12] G.-H. Gweon, J. D. Denlinger, J. A. Clack, J. W. Allen, C. G. Olson, E. DiMasi, M. C. Aronson, B. Foran, and S. Lee, Direct observation of complete Fermi surface, imperfect nesting, and gap anisotropy in the high-temperature incommensurate charge-density-wave Compound $SmTe_3$, Phys. Rev. Lett. **81**, 886 (1998).

[13] V. Brouet, W. L. Yang, X. J. Zhou, Z. Hussain, N. Ru, K. Y. Shin, I. R. Fisher, and Z. X. Shen, Fermi surface reconstruction in the CDW state of $CeTe_3$ observed by photoemission, Phys. Rev. Lett. **93**, 126405 (2004).

[14] K. Yumigeta, Y. Qin, H. Li, M. Blei, Y. Attarde, C. Kopas, and S. Tongay, Advances in rare-earth tritelluride quantum materials: Structure, properties, and synthesis, Adv. Sci. (Weinh.) **8**, e2004762 (2021).

[15] N. Ru, J.-H. Chu, and I. R. Fisher, Magnetic properties of the charge density wave compounds $RTe_3$ (R=Y, La, Ce, Pr, Nd, Sm, Gd, Tb, Dy, Ho, Er, and Tm), Phys. Rev. B **78**, 012410 (2008).

[16] J. J. Hamlin, D. A. Zocco, T. A. Sayles, M. B. Maple, J.-H. Chu, and I. R. Fisher, Pressure-induced superconducting phase in the charge-density-wave compound terbium tritelluride, Phys. Rev. Lett. **102**, 177002 (2009).

[17] D. A. Zocco, J. J. Hamlin, K. Grube, J.-H. Chu, H.-H. Kuo, I. R. Fisher, and M. B. Maple, Pressure dependence of the charge-density-wave and superconducting states in $GdTe_3$, $TbTe_3$, and $DyTe_3$, Phys. Rev. B **91**, 205114 (2015).

[18] S. Akatsuka, S. Esser, S. Okumura, R. Yamabe, R. Yamada, M. M. Hirschmann, S. Aji, J. S. White, S. Gao, Y. Onuki, T.-H. Arima, T. Nakajima, and M. Hirschberger, Non-coplanar helimagnetism in the layered van-der-Waals metal $DyTe_3$, Nat. Commun. **15**, 4291 (2024).

[19] D. Wu, Q. M. Liu, S. L. Chen, G. Y. Zhong, J. Su, L. Y. Shi, L. Tong, G. Xu, P. Gao, and N. L. Wang, Layered semiconductor $EuTe_4$ with charge density wave order in square tellurium sheets, Phys. Rev. Mater. **3**, 024002 (2019).

[20] Q. Q. Zhang, Y. Shi, K. Y. Zhai, W. X. Zhao, X. Du, J. S. Zhou, X. Gu, R. Z. Xu, Y. D. Li, Y. F. Guo, Z. K. Liu, C. Chen, S.-K. Mo, T. K. Kim, C. Cacho, J. W. Yu, W. Li, Y. L. Chen, J.-H. Chu, and L. X. Yang, Thermal hysteretic behavior and negative magnetoresistance in the charge density wave material $EuTe_4$, Phys. Rev. B **107**, 115141 (2023).

[21] C. Zhang, Q.-Y. Wu, Y.H. Yuan, X. Zhang, H. Liu, Z.-T. Liu, H.-Y. Zhang, J.-J. Song, Y.-Z Zhao, F.-Y Wu, S.-Y Liu, B. Chen, X.-Q. Ye, S.-T. Cui, Z. Sun, X.-F. Tang, J. He, H.-Y. Liu, Y.-X. Duan, Y.-F. Guo, and J.-Q. Meng, Angle-resolved photoemission spectroscopy

study of charge density wave order in the layered semiconductor EuTe$_4$, Phys. Rev. B. **106**, L201108 (2022).

[22] B. Q. Lv, A. Zong, D. Wu, Z. Nie, Y. Su, D. Choi, B. Ilyas, B. T. Fichera, J. Li, E. Baldini, M. Mogi, Y.-B. Huang, H. C. Po, S. Meng, Y. Wang, N. L..Wang, and N. Gedik, Coexistence of interacting charge density waves in a layered semiconductor, Phys. Rev. Lett. **132**, 206401 (2024).

[23] B. Q. Lv, A. Zong, D. Wu, A. V. Rozhkov, B. V. Fine, S.-D. Chen, M. Hashimoto, D.-H. Lu, M. Li, Y.-B. Huang, J. P. C. Ruff, D. A. Walko, Z. H. Chen, I. Hwang, Y. Su, X. Shen, X. Wang, F. Han, H. C. Po, Y. Wang, P. J. Herrero, X. Wang, H. Zhou, C.-J. Sun, H. Wen, Z.-X. Shen, N. L. Wang, and N. Gedik, Unconventional hysteretic transition in a charge density wave, Phys. Rev. Lett. **128**, 036401 (2022).

[24] R. Rathore, A. Pathak, M. K. Gupta, R. Mittal, R. Kulkarni, A. Thamizhavel, H. Singhal, A. H. Said, and D. Bansal, Evolution of static charge density wave order, amplitude mode dynamics, and suppression of Kohn anomalies at the hysteretic transition in EuTe$_4$, Phys. Rev. B. **107**, 024101 (2023).

[25] K. Xiao, W.-H. Dong, X. Wang, J. Yu, D. Fu, Z. Hu, Y. Guo, Q. Zhang, X. Hou, Y. Guo, L. Yang, Y. Xu, P. Tang, W. Duan, Q. Xue, and W. Li, Hidden Charge Order and Multiple Electronic Instabilities in EuTe$_4$, Nano Lett. **24**, 7681 (2024).

[26] B. Q. Lv, Y. Su, A. Zong, Q. Liu, D. Wu, N. F. Q. Yuan, Z. Nie, J. Li, S. Sarker, S. Meng, J. P. C. Ruff, N. L. Wang, and N. Gedik, Large Moiré Superstructure of Stacked Incommensurate Charge Density Waves, http://arxiv.org/abs/2501.09715.

[27] Q. Liu, D. Wu, T. Wu, S. Han, Y. Peng, Z. Yuan, Y. Cheng, B. Li, T. Hu, L. Yue, S. Xu, R. Ding, M. Lu, R. Li, S. Zhang, B. Lv, A. Zong, Y. Su, N. Gedik, Z. Yin, T. Dong, and N. Wang, Room-temperature non-volatile optical manipulation of polar order in a charge density wave, Nat. Commun. **15**, 8937 (2024).

[28] R. Venturini, M. Rupnik, J. Gašperlin, J. Lipič, P. Šutar, Y. Vaskivskyi, F. Ščepanović, D. Grabnar, D. Golež, and D. Mihailovic, Electrically Driven Non-Volatile Resistance Switching between Charge Density Wave States at Room Temperature, http://arxiv.org/abs/2412.13094.

[29] H. Takahashi, K. Aono, Y. Nambu, R. Kiyanagi, T. Nomoto, M. Sakano, K. Ishizaka, R. Arita, and S. Ishiwata, Competing spin modulations in the magnetically frustrated semimetal EuCuSb, Phys. Rev. B. **102**, 174425 (2020).

[30] H. Takahashi, K. Akiba, M. Takahashi, A. H. Mayo, M. Ochi, T. C. Kobayashi, and S. Ishiwata, Superconductivity in a Magnetic Rashba Semimetal EuAuBi, J. Phys. Soc. Jpn. **92**, 013701 (2023).

[31] See the Supplemental Material at /// for the results of the SEM, EDX, X-ray diffraction, TG-DTA, magnetoresistance, and Hall resistivity, and the analysis for the magnetic susceptibility, and the estimation of the power factor, and the details of the band calculations. The Supplemental Material includes Refs. [19-23,37,39,52-55]

[32] P. Giannozzi, S. Baroni, N. Bonini, M. Calandra, R. Car, C. Cavazzoni, D. Ceresoli, G. L Chiarotti, M. Cococcioni, I. Dabo, A. D. Corso, S. de Gironcoli, S. Fabris, G. Fratesi, R. Gebauer, U. Gerstmann, C. Gougoussis, A. Kokalj, M. Lazzeri, L. M. Samos, N. Marzari, F. Mauri, R. Mazzarello, S. Paolini, A. Pasquarello, L. Paulatto, C. Sbraccia, S. Scandolo, G. Sclauzero, A. P Seitsonen, A. Smogunov, P. Umari, and R. M. Wentzcovitch, QUANTUM ESPRESSO: a modular and open-source software project for quantum simulations of materials, J. Phys. Condens. Matter **21**, 395502 (2009).

[33] P. Giannozzi, O. Andreussi, T. Brumme, O. Bunau, M. B. Nardelli, M. Calandra, R. Car, C. Cavazzoni, D. Ceresoli, M. Cococcioni, N. Colonna, I. Carnimeo, A. D. Corso, S. de Gironcoli, P. Delugas, R. A. DiStasio Jr, A. Ferretti, A. Floris, G. Fratesi, G. Fugallo, R. Gebauer, U. Gerstmann, F. Giustino, T. Gorni, J. Jia, M. Kawamura, H.-Y. Ko, A. Kokalj, E. Küçükbenli, M. Lazzeri, M. Marsili, N. Marzari, F. Mauri, N. L. Nguyen, H.-

V. Nguyen, A. Otero-de-la-Roza, L. Paulatto, S. Poncé, D. Rocca, R. Sabatini, B. Santra, M. Schlipf, A. P. Seitsonen, A. Smogunov, I. Timrov, T. Thonhauser, P. Umari, N. Vast, X. Wu, and S. Baroni, Advanced capabilities for materials modelling with Quantum ESPRESSO, J. Phys. Condens. Matter **29**, 465901 (2017).

[34] *Quantum Espresso*, http://www.quantum-espresso.org.

[35] A. Dal Corso, Pseudopotentials periodic table: From H to Pu, Comput. Mater. Sci. **95**, 337 (2014).

[36] M. Kawamura, FermiSurfer: Fermi-surface viewer providing multiple representation schemes, Comput. Phys. Commun. **239**, 197 (2019).

[37] A. Pathak, M. K. Gupta, R. Mittal, and D. Bansal, Orbital- and atom-dependent linear dispersion across the Fermi level induces charge density wave instability in $EuTe_4$, Phys. Rev. B. **105**, 035120 (2022).

[38] G. K. H. Madsen and D. J. Singh, BoltzTraP. A code for calculating band-structure dependent quantities, Comput. Phys. Commun. **175**, 67 (2006).

[39] H. Ghamri, Y. Djaballah, and A. Belgacem-Bouzida, Thermodynamic modeling of the Eu–Te and Te–Yb systems, J. Alloys Compd. **643**, 121 (2015).

[40] J. Yin, C. Wu, L. Li, J. Yu, H. Sun, B. Shen, B. A. Frandsen, D.-X. Yao, and M. Wang, Large negative magnetoresistance in the antiferromagnetic rare-earth dichalcogenide $EuTe_2$, Phys. Rev. Mater. **4**, 013405 (2020).

[41] H. Takahashi, Y. Yasui, I. Terasaki, and M. Sato, Effects of ppm-level imperfection on the transport properties of $FeSb_2$ single crystals, J. Phys. Soc. Jpn. **80**, 054708 (2011).

[42] B. C. Sales, D. Mandrus, and R. K. Williams, Filled skutterudite antimonides: A new class of thermoelectric materials, Science **272**, 1325 (1996).

[43] D.-L. Ma, P. Bag, Y.-K. Kuo, C.-N. Kuo, and C. S. Lue, Thermoelectric and specific heat characteristics of the charge density wave compounds $RTe_3$ ($R$ = Gd, Tb, Dy, Ho, and Er), Phys. Rev. B. **111**, 165420 (2025).

[44] H. Imai, Y. Shimakawa, and Y. Kubo, Large thermoelectric power factor in $TiS_2$ crystal with nearly stoichiometric composition, Phys. Rev. B **64**, 241104(R) (2001).

[45] T. Caillat, A. Borshchevsky, and J.-P. Fleurial, Properties of single crystalline semiconducting $CoSb_3$, J. Appl. Phys. **80**, 4442 (1996).

[46] T. H. Geballe and G. W. Hull, Seebeck effect in germanium, Phys. Rev. **94**, 1134 (1954).

[47] C. Collignon, P. Bourges, B. Fauqué, and K. Behnia, Heavy nondegenerate electrons in doped strontium titanate, Phys. Rev. X. **10**, 031025 (2020).

[48] V. Zlatić, R. Monnier, J. K. Freericks, and K. W. Becker, Relationship between the thermopower and entropy of strongly correlated electron systems, Phys. Rev. B **76**, 085122 (2007).

[49] J. Mravlje and A. Georges, Thermopower and entropy: Lessons from $Sr_2RuO_4$, Phys. Rev. Lett. **117**, 036401 (2016).

[50] H. Takahashi, S. Ishiwata, R. Okazaki, Y. Yasui, and I. Terasaki, Enhanced thermopower via spin-state modification, Phys. Rev. B **98**, 024405 (2018).

[51] T. D. Yamamoto, H. Taniguchi, Y. Yasui, S. Iguchi, T. Sasaki, and I. Terasaki, Magneto-thermopower in the Weak Ferromagnetic Oxide $CaRu_{0.8}Sc_{0.2}O_3$: An Experimental Test for the Kelvin Formula in a Magnetic Material, Journal of the Physical Society of Japan **86**, 104707 (2017).

[52] H. Sakai, K. Ikeura, M. S. Bahramy, N. Ogawa, D. Hashizume, J. Fujioka, Y. Tokura, and S. Ishiwata, Critical enhancement of thermopower in a chemically tuned polar semimetal $MoTe_2$, Sci. Adv. **2**, e1601378 (2016).

[53] H. Takahashi, T. Akiba, K. Imura, T. Shiino, K. Deguchi, N. K. Sato, H. Sakai, M. S. Bahramy, and S. Ishiwata, Anticorrelation between polar lattice instability and superconductivity in the Weyl semimetal candidate $MoTe_2$, Phys. Rev. B **95**, 100501 (2017).

[54] H. Takahashi, K. Hasegawa, T. Akiba, H. Sakai, M. S. Bahramy, and S. Ishiwata, Giant enhancement of cryogenic thermopower by polar structural instability in the pressurized

semimetal MoTe$_2$, Phys. Rev. B. **100**, 195130 (2019).

[55] A. Nakano, A. Yamakage, U. Maruoka, H. Taniguchi, Y. Yasui, and I. Terasaki, Giant Peltier conductivity in an uncompensated semimetal Ta$_2$PdSe$_6$, J. Phys. Energy **3**, 044004 (2021).

[56] R. Rathore, H. Singhal, V. Dwij, M. K Gupta, A. Pathak, J. A. Chakera, R. Mittal, A. P. Roy, A. Babu, R. Kulkarni, A. Thamizhavel, A. H. Said, and D. Bansal, Nonlocal probing of amplitude mode dynamics in charge-density-wave phase of EuTe$_4$, Ultrafast Sci **3**, 0041 (2023).

[57] L. Huang, T. Liu, A. Huang, G. Yuan, J. Wang, J. Liao, X. Lei, Q. Zhang, and Z. Ren, Enhanced thermoelectric performance of nominal 19-electron half-Heusler compound NbCoSb with intrinsic Nb and Sb vacancies, Mater. Today Phys. **20**, 100450 (2021).

[58] M. Mitra, A. Benton, M. S. Akhanda, J. Qi, M. Zebarjadi, D. J. Singh, and S. J. Poon, Conventional Half-Heusler alloys advance state-of-the-art thermoelectric properties, Mater. Today Phys. **28**, 100900 (2022).

[59] Y. Ren, Q. Jiang, J. Yang, Y. Luo, D. Zhang, Y. Cheng, and Z. Zhou, Enhanced thermoelectric performance of MnTe via Cu doping with optimized carrier concentration, J. Mater. **2**, 172 (2016).

[60] Y. Zheng, T. Lu, Md M. H. Polash, M. Rasoulianboroujeni, N. Liu, M. E. Manley, Y. Deng, P. J. Sun, X. L. Chen, R. P. Hermann, D. Vashaee, J. P. Heremans, and H. Zhao, Paramagnon drag in high thermoelectric figure of merit Li-doped MnTe, Sci. Adv. **5**, eaat9461 (2019).

[61] H. Fukuyama and M. Ogata, Theory of phason drag effect on thermoelectricity, Phys. Rev. B. **102**, 205136 (2020).